\documentclass[conference]{IEEEtran}

\usepackage{todonotes}
\usepackage{adjustbox}
\usepackage{graphicx}
\usetikzlibrary{calc}
\usepackage{booktabs}
\usepackage{paralist}
\usepackage{amssymb}
\usepackage{amsmath}
\usepackage{centernot}
\usepackage[sort]{cite}
\usepackage{diagbox}
\usepackage{caption,subcaption}

\newcommand{\comment}[1]{}

\newcommand{\NotEmpty}{ \mathrel{\ooalign{\hss$\Diamond$\hss\cr
  \kern-1pt\raise0ex\hbox{\scalebox{1}{$\not$}}}}}

\usepackage[printonlyused]{acronym}
\acrodef{PCN}{Payment Channel Network}
\acrodef{P3}{Privacy-Preserving Payment}
\newcommand{\PPPSystem}{\ac{P3} system} 
\newcommand{\PPPSystems}{\ac{P3} systems} 
\renewcommand{\paragraph}[1]{
    \noindent \textit{#1}.}
    
\newif\iflong
\longtrue

\acrodef{PPT}{Probabilistic Polynomial Time}

\begin{document}

\title{The Danger of Small Anonymity Sets in Privacy-Preserving Payment Systems}

\author{
\IEEEauthorblockN{Christiane Kuhn}
\IEEEauthorblockA{Karlsruhe Institute of Technology\\
christiane.kuhn@kit.edu}
\and
\IEEEauthorblockN{Aniket Kate}
\IEEEauthorblockA{Purdue University\\
aniket@purdue.edu}
\and
\IEEEauthorblockN{Thorsten Strufe}
\IEEEauthorblockA{Karlsruhe Institute of Technology\\
thorsten.strufe@kit.edu}
}

\maketitle

\begin{abstract}
Unlike suggested during their early years of existence, Bitcoin and similar cryptocurrencies in
fact offer significantly less privacy as compared to traditional banking.   
A myriad of privacy-enhancing extensions to those cryptocurrencies as well as several clean-slate privacy-protecting
cryptocurrencies have been proposed in turn.

To convey a better understanding of the protection of popular design decisions, we investigate expected anonymity set sizes in an initial simulation study. 
The large variation of expected transaction values yields soberingly small effective anonymity sets for protocols that leak transaction values. 
We hence examine the effect of preliminary, intuitive strategies for merging groups of payments into larger anonymity sets, for instance by choosing from pre-specified value classes.
The results hold promise, as they indeed induce larger anonymity sets at comparatively low cost, depending on the corresponding strategy.
\end{abstract}

\section{Introduction}\label{sec:intro}

Experience with Bitcoin, once cherished as the privacy-preserving alternative to other digital payment methods, has demonstrated that it effectively only provides very weak pseudonymity for transactions~\cite{bitiodine,imc-meiklejohn,ron-shamir,ripple-pets,chainalysis,Koshy2014,Reid2013}.
\PPPSystems{} have thus become the focus of interest, as the fundamental requirement for anonymous payments has not lost its relevance.

To shed first light on the \emph{real world} implications of popular design decisions,
we conduct a simulation study.
We choose four representative cases for sender protection and compare the resulting anonymity set sizes. 
We first generate various sets of payment data using a model that is based on measured assumptions, and later employ a real world data set from Ripple. 
The results allow us to review the effect of typical design decisions, like leaking the value of payments or deciding on delays, on the effective anonymity set size and its potential reduction.
Further, we identify potential adaptations: we observe effects of changing system parameters, like increasing latency or payment frequency. We also provide evidence for improvements by simple strategies: scaled value buckets, for instance, prove useful to considerably increase the anonymity sets, at the cost of paying at most 10\% more than the intended value.

In summary, our main contributions are:
\begin{compactitem}[--]
    \item experimental study relating design decisions to expected anonymity set sizes, on synthetic and real world data, and
    \item identification and early investigations of improvement opportunities for future approaches.
\end{compactitem}


\section{Background}\label{sec:background}
\subsection{Privacy Goals}

For this work, we will focus on sender protection. More precisely we understand the protection as different kinds of \emph{Sender Anonymity}. Following the work of Pfitzmann and Hansen \cite{pfitzman10terminology}, \emph{Anonymity} of a subject describes the state of being not identifiable within a set of subjects. This set is the \emph{Anonymity Set}. As we discuss below, different anonymity sets are conceivable as depending on the employed protection measures different users contribute to the anonymity of a single sender.

\subsection{Sources of Payment Data and Protocols}

\emph{Bitcoin}~\cite{nakamoto2019bitcoin} is the most known real-world cryptocurrency that stores transactions publicly on a blockchain. While bitcoin decentralizes the payment execution, it only offers a very weak form of pseudonymous privacy and users can be identified by analyzing the payment graph in detail. 

\emph{Ethereum}~\cite{ethereum} introduces Ether as alternative cryptocurrency, but does not aim to provide privacy. We however use its 1.2 million transactions per day as one of our baselines for the usage frequency of cryptocurrencies.

\emph{Ripple}~\cite{armknecht2015ripple} and \emph{Stellar}~\cite{stellar} are IOweYou credit networks~\cite{fugger2004money,formation-credit-networks}. In a credit network participants allow other participants a credit based on the trust between the entities. This results in a graph where vertices represent participants and edge weights the allowed credit. Similar to Bitcoin, a publicly available global ledger records the transactions and is updated upon consensus.

Somewhat similar to credit networks, in \emph{Payment Channel Networks (PCNs)} participants lock funds on the blockchain via smart contracts to then be able to do multiple payments before closing the channel and updating the blockchain accordingly. This aggregation of transactions naturally introduces some privacy protection.

 \section{Considered Anonymity Sets}
 We consider the following typical anonymity design decisions for \PPPSystems{}. The anonymity set consists of:

\subsubsection{All Users (all)}
 Any user that participates in the protocol cannot be distinguished from the real sender and the anonymity set includes \emph{all users} of the \PPPSystem{}. Note that the \PPPSystem{} therefore has to employ protection measures that even hide the activity of each user from the adversary.

\subsubsection{Active Senders (active)}
Any user that sent something during the observed payments cannot be distinguished from the real sender and the anonymity set includes \emph{all active senders}. Note that the \PPPSystem{} therefore has to employ protection measures that even hide the amount that the sender transfers from the adversary.

\subsubsection{Active Senders with the Same Value (active + value)}
The sender anonymity set hence differs just due to the fixed value; it thus includes \emph{all active senders that send the same value}. Note that the \PPPSystem{} therefore has to employ protection measures that conceals the senders regardless of their position in the topology.

\subsubsection{Senders whose Payments meet on their Path trough the \PPPSystems{} (path)}
For PCNs naturally only transactions that "meet" during processing mix with each other and contribute to the same anonymity set. More precisely, we require that the payments meet at an honest node on the path through the PCN. Thus, any sender with whom a transaction can be exchanged this way and leads to a plausible routing path is indistinguishable to the real sender for an adversary. 
For our simulation, we are choosing a routing algorithm that does not include any loops and hence consider only alternatives without a loop in the routing path. (We show the effect of including loops 
in Appendix~\ref{app:loops}.)

\begin{figure}
\centering
\includegraphics[width=0.48\textwidth]{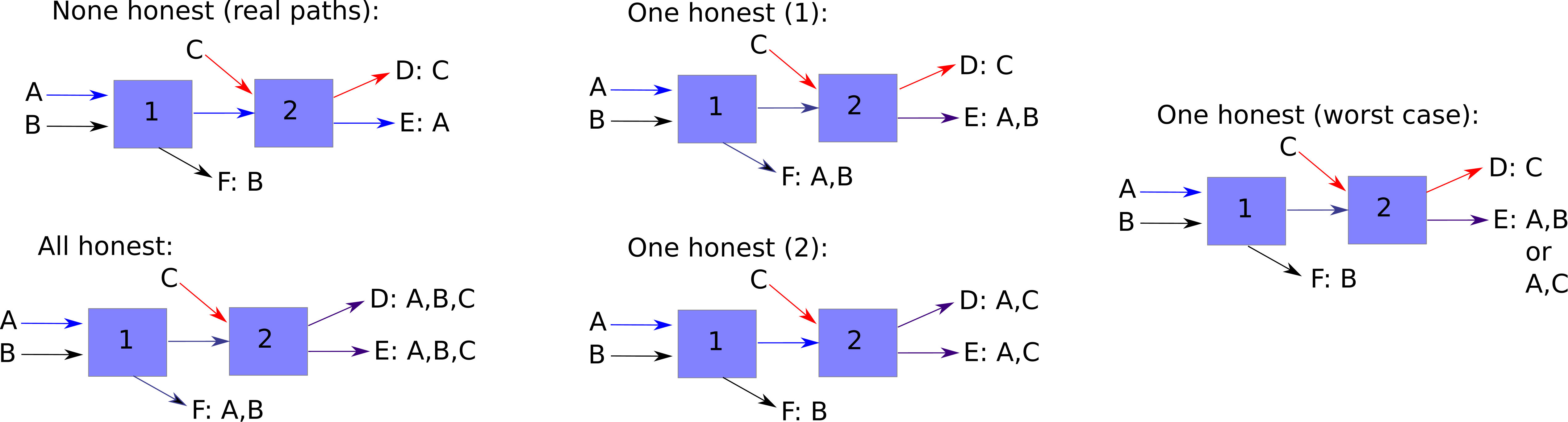}
\caption{Maximal (all nodes honest) and minimal anonymity sets (worst case node per path is honest) illustrated.}
\label{fig:MinMaxAnonSets}
\end{figure}

Further, these anonymity sets depend on the adversary model as we consider only paths with a common \emph{honest} node. We will be looking at the two (non-trivial) extreme cases below (as illustrated in Figure~\ref{fig:MinMaxAnonSets}): (i) all nodes are honest and (ii) exactly one node per payment path is honest. Thus the first gives us the maximal anonymity set reachable for this setting. For the second case, we decide on the worst intermediate node, i.e. the one with the least other payments, to be honest. This is the minimal anonymity set (except for the unprotected case when there is no honest node on the path, and the anonymity set trivially only includes the real sender).

\section{Anonymity Sets for Realistic User Behavior}

In this section we perform an initial investigation of the anonymity set sizes that are actually to be expected in practice.
As they rely on specific payment choices, we  model realistic user behavior and simulate actual payments accordingly for our analysis.
It indeed is hard to decide on realistic parameters, especially if sender-receiver pairs, or a specific payment network topology are needed - as for instance in path-based anonymity sets.
We hence first study anonymity sets that we can determine without such information, and later study an example including path-based anonymity sets on real data.

{Considering the anonymity sets that are to be expected in reality to be prohibitively small in several cases, we also perform an initial exploration study of preliminary strategies that may help increase the anonymity set sizes at limited cost, in addition.}
Intuitive approaches are to increase the transaction value to the next higher known to be used value, delaying transactions to increase the chance for some additional payments with identical value to be created, or to generally restrict all transaction values to specific, allowed value classes.

\subsection{Setting and Data}\label{Sec:SimSettingGenerated}
\subsubsection{Setting} To identify the impact of deciding on a different protection, we calculate anonymity set sizes for each version with a round based python simulation.
All payments that are initiated within an epoch time slot are considered to be mixed with each other for the \emph{active} or \emph{value anonymity set}. 
This models e.g. a tumbler with fixed times to output payments again. 
We expressly have not implement the tumbler (or any other \PPPSystem{}), but merely draw payment events according to the user model and count the number of distinct users with the considered criteria in each epoch, assuming that they will choose the same tumbler (or corresponding anonymization component).

\subsubsection{Data}
We report on the results based on a group size of 100~000 users for our experiments, performing 30 repetition for each parameter set. 
Repeating the experiments with 10~000 and 1~000~000 users confirmed our findings.

We modeled our assumptions based on \cite{moser2017price}, which measured JoinMarket, a trading platform for privacy enhanced payments with Bitcoin. 
The values consequently are drawn following a lognormal distribution with mean of 84 USD and standard deviation of $2.4$. Thus, the quantiles are ~16 USD (25\%), 84 USD (50\%) and ~420 USD (75\%).
We rounded all values to full USD to improve the chances for the value to match an existing anonymity set, and the resulting transactions cover values from one to several million USD as in the published data.

Information about senders is sparse as the networks try to protect the senders' anonymity. We thus use the best guess we can make for sending behavior: inter-sending times that are distributed according to a Poisson distribution. To account for the uncertainty, we vary the parameter $\lambda$ ($ 10, 50, 100$) and to give them a real world meaning, we match $\lambda=50$ to real world time  according to the total payments made per day in different networks as shown in Table~\ref{tab:timePerTick}. So $\lambda =50$ represents the current payment situation, $\lambda=10$  the situation that the payment networks are used much more frequently and $\lambda=100$ the situation that the networks are used less frequent.
We ensure that the simulations reach a stable state before we measure the  anonymity sets.

\begin{table}[h!]
  \begin{center}
    \caption{Real world time per simulation time unit (assuming different usage frequencies)}
    \label{tab:timePerTick}
    \resizebox{\columnwidth}{!}{
    \begin{tabular}{ c |c |c| c} 
      \textbf{Usage frequency} & \multicolumn{2}{c}{\textbf{Time per simulation time unit}} & \textbf{Payments per day}\\
      &\textbf{Generated Data} & \textbf{Ripple Data} \\\toprule
      Ethereum~\cite{ethereum} & ~2.5 min & 3.8 sec& 1.2 mio\\
      Bitcoin~\cite{nakamoto2019bitcoin} & ~9.5 min & 15 sec &0.3 mio\\
      Germany (incl. cash) &  1.7 sec & 45 millisec &99.6 mio\\
      Canada (incl. cash) & 7.5 sec & ~200 millisec & 22.7 mio\\ \bottomrule
    \end{tabular}
    }
  \end{center}
\end{table}

\subsection{Results}
\subsubsection{Anonymity Set Comparison}
Figure~\ref{fig:AnonSetsGen} and Table~\ref{tab:AnonSetsGen} show an overview of the distributions of anonymity set sizes. 
Increasing either epoch length or payment frequency 
results in growing anonymity sets, as we would expect. 

Comparing the effective size of anonymity sets with matching values to the overall number of users (100 000) yields sobering results. The same holds for the number of simultaneously active users given expected payment frequencies and epoch times at or below 25 minutes. Both decrease when epoch times or payment frequencies are reduced, as expected.

The anonymity set sizes for the value-based sets clearly are unacceptably small, with about 3~000 entirely deanonymized payments on average. 
The number of active senders remains on the order of tens of thousands of users, which may be considered acceptable as potential anonymity set size.
We hence investigated the effect on the anonymity sets that an adaption of the transaction values or payment delay has.

\begin{table}[htb]
  \begin{center}
    \caption{Lower quartile, median and upper quartile for anonymity set sizes under varying frequencies and epochs (active sizes rounded)}
    \label{tab:AnonSetsGen}
\resizebox{0.95\linewidth}{!}{
\begin{tabular}{|l|c|c|c|}\hline
\diagbox[width=8em]{\textbf{Frequency}}{\textbf{Epoch}}&
  \textbf{12.5 min} & \textbf{25 min} & \textbf{125 min} \\ \hline
 \textbf{high:} active&  45.2k \textbf{45.3k} 45.4k & 83.6k \textbf{83.7k} 83.8k& 100k \textbf{100k} 100k
 \\
active+value &  1 \textbf{1} 3 & 1 \textbf{1} 4& 1 \textbf{1} 4 \\ 
\hline
\textbf{normal:} active&  10k \textbf{10k} 10.1k & 20.2k \textbf{20.3k} 20.3k& 93.5k \textbf{93.6k} 93.6k  \\
active+value & 1 \textbf{1} 3 & 1 \textbf{1} 3&1 \textbf{1} 4  \\ 
\hline
\textbf{low:} active&  6.2k \textbf{6.2k} 6.3k & 12.6k \textbf{12.6k} 12.7k & 54.9k \textbf{55k} 55k  \\
active+value &  1 \textbf{1} 2 & 1 \textbf{1} 3& 1 \textbf{1} 3 \\ 
\hline
\end{tabular}}
\end{center}
\end{table}

\subsubsection{Paying more against deanonymization}
If a user would know all payment values of an epoch and decide to increase their payment to reach the next higher value payed during the same epoch, she could create a value based anonymity set of at least 2. The distribution of relative cost over all additional costs $>0$ is shown in Figure~\ref{fig:payMoreGen}.
The additional cost to ensure that at least two users share the same value in most cases ranges around a few to a few dozen percent of the original value. 
Since very high payment values do occur with small probability, there exist a few very costly outliers.

\iflong

\begin{figure*}
\centering
\includegraphics[width=0.9\textwidth]{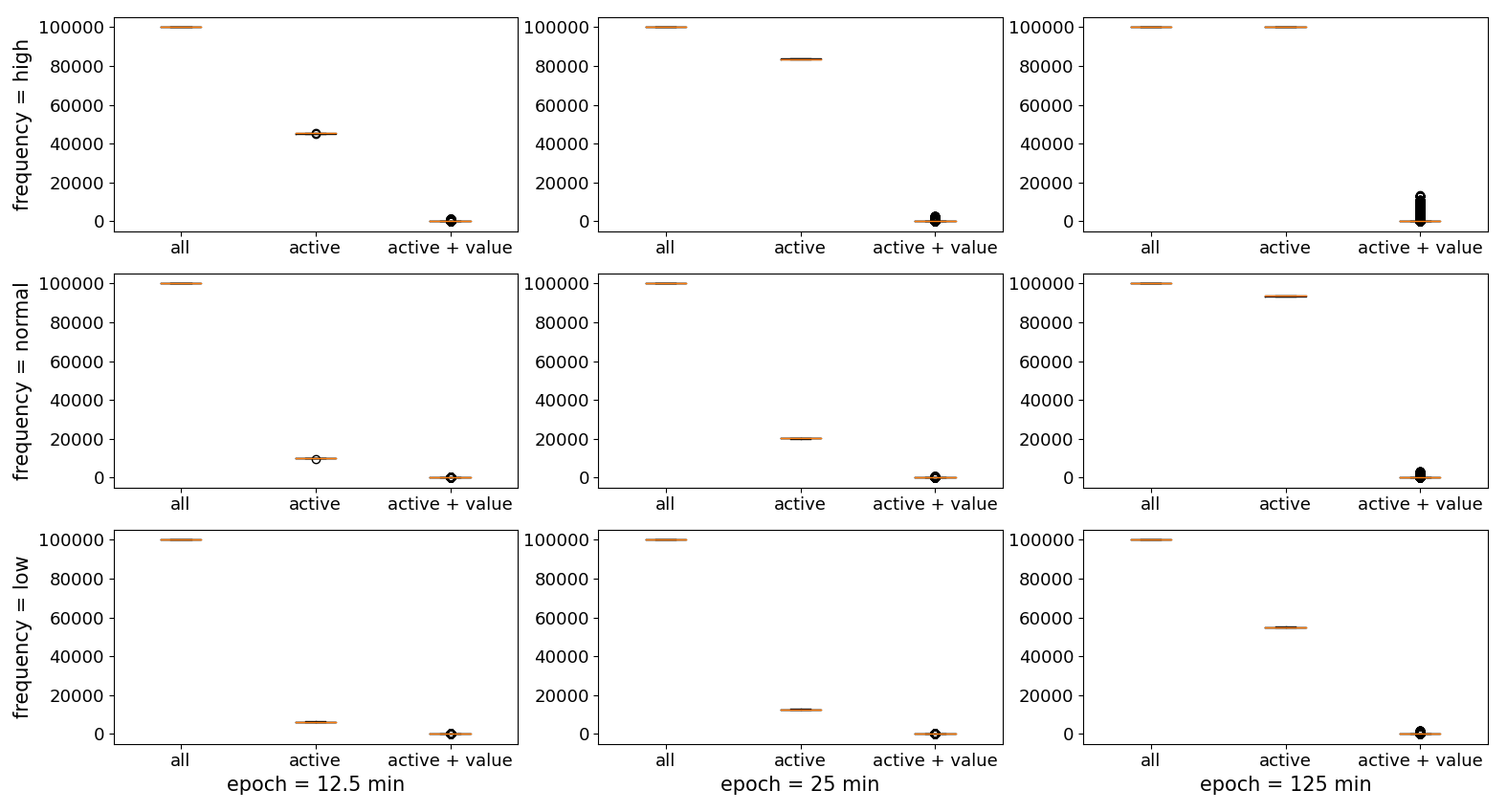}
\caption{Anonymity set sizes for the generated data. Unless a very high paying frequency or waiting time (~ 2 hours) is assumed, the set of active senders is drastically smaller than the set of all users. Even in the best case (high frequency and long epochs) the number of active users with the same value is close to 1.}
\label{fig:AnonSetsGen}
\end{figure*}

\begin{figure}
\centering
\includegraphics[width=0.48\textwidth]{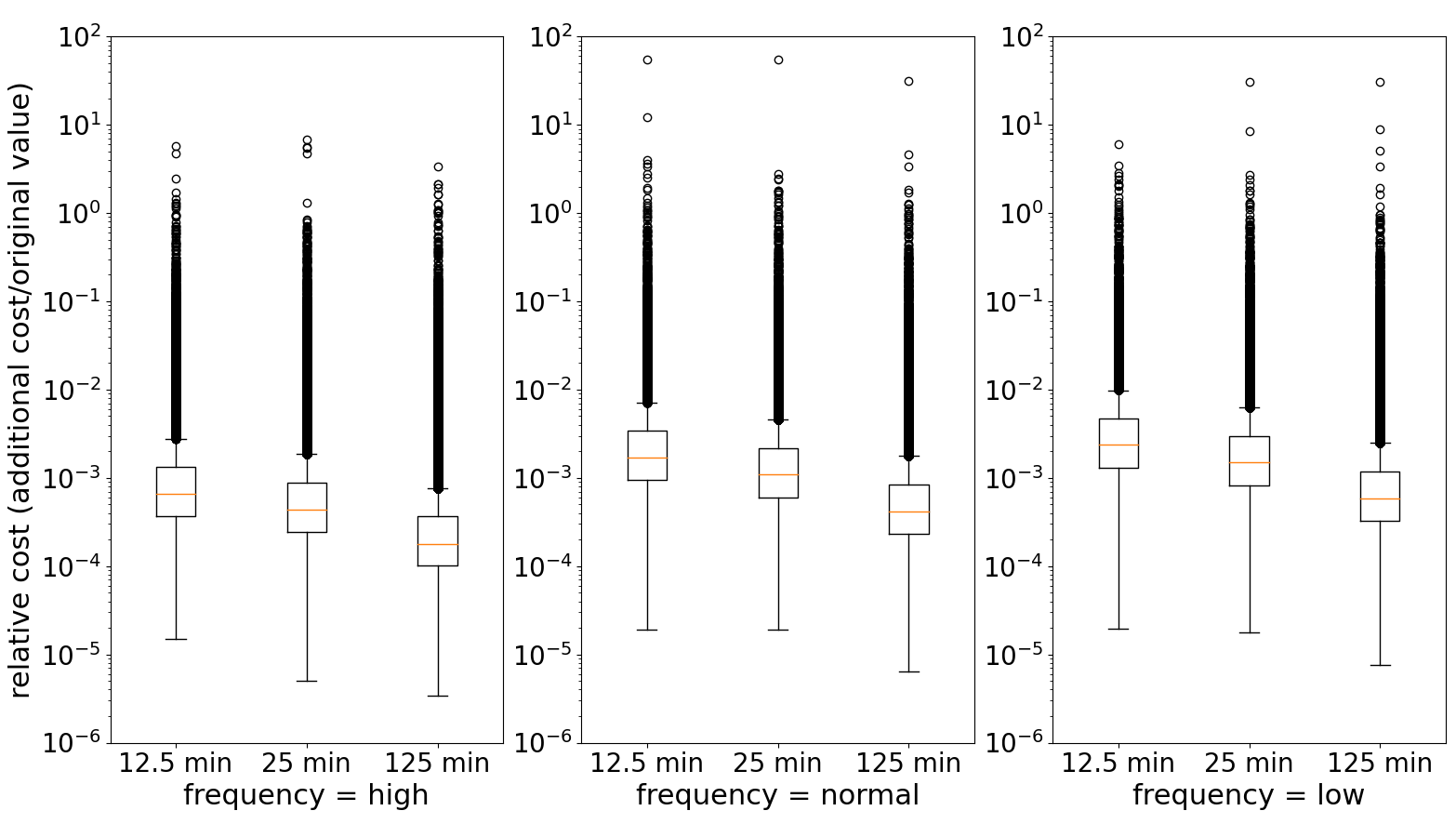}
\caption{Relative cost to not be unique, i.e. $\frac {\text{additional cost}}{\text{desired value}}$ (other frequencies introduce only slightly different costs.)}
\label{fig:payMoreGen}
\end{figure}

\begin{figure}
\centering
\includegraphics[width=0.47\textwidth] {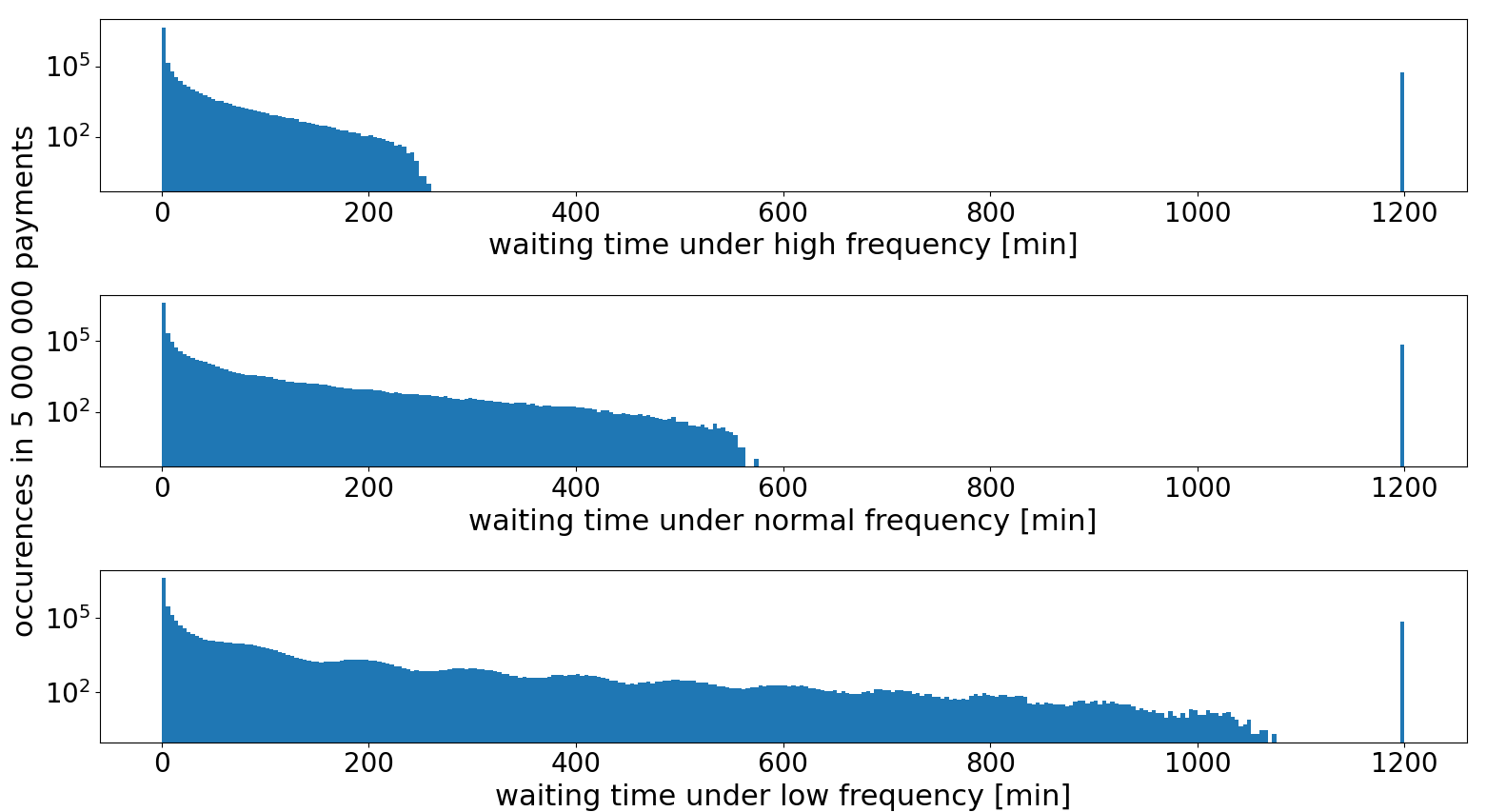}
\caption{Waiting time until another payment with the same value occurs investigated for 5 million generated payments. 1200 represents that no such payment occurred in the generated payment setting. For the high frequency waiting times up to 250 minutes and under low frequency up to 1100 minutes occur.}
\label{fig:waitLongerGen}
\end{figure}

\begin{figure*}
\centering
\includegraphics[width=0.8\textwidth]{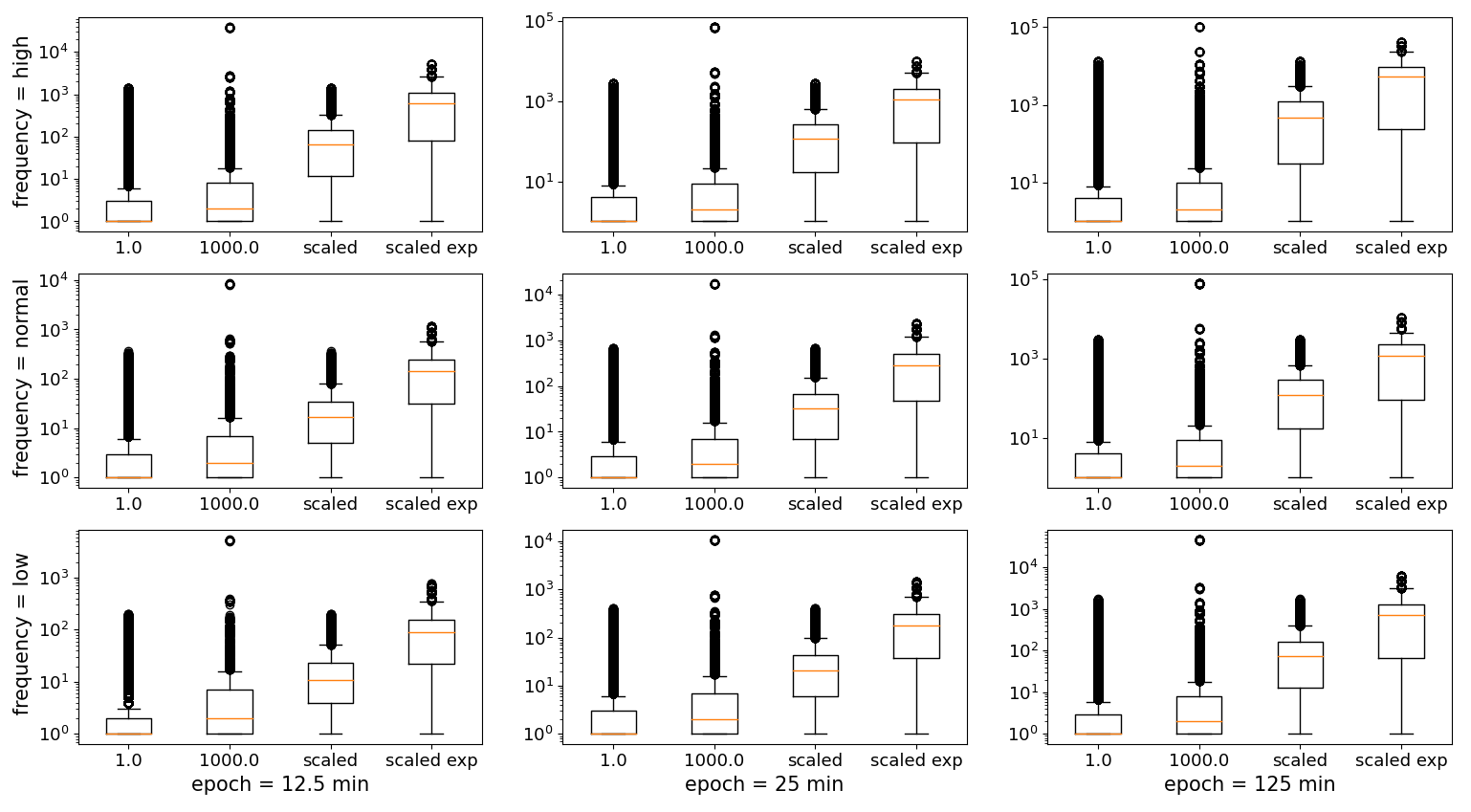}
\caption{Anonymity set sizes with value buckets. The results demonstrate a great improvement for all epoch times and frequencies when using the scaled buckets that at the same time also cause lower relative cost than fixed bucket sizes.}
\label{fig:valueBucketsGen}
\end{figure*}

\fi

\subsubsection{Waiting longer against deanonymization}
With perfect knowledge of all payments, another strategy could be to delay a transaction in order to find an anonymity set of at least size 2.
We report the necessary waiting time per user in Fig.~\ref{fig:waitLongerGen}.
The time to wait depends highly on the usage frequency.
But even with the highest usage frequency it remains over 2 hours for a substantial part of the payments.

\subsubsection{Introducing Value Buckets}
Consider all users to participate in a cooperative strategy:
Reducing the number of possible payment values to choose from increases the chance of collisions, and hence the expected anonymity set size.

A simple approach is to use fixed value buckets, for instance to round the values up to full 10-, 100-, or even 1~000-dollar amounts. In the last case of course a user wanting to buy something for 1 dollar would need to pay 1~000. 
Another option is to reduce the relative additional incurred cost, i.e. the amount of money a user has to pay divided by the intended value. 
We hence investigate two types of scaling buckets: 1) a cheap variant that limits the relative additional cost to 10\% of the original value and 2) a more expensive variant that limits the relative additional cost to 100\% of the original value, but still achieves a much better relative cost than the above strategy of rounding to full USD 1~000 amounts.

Our cheap scaling buckets assume the new value ($v_{cheap}$) to be calculated from the original value ($v_{ori}$) as follows:
\[v_{cheap}=\left \lceil{
\frac{v_{ori}}
{10^{\left \lfloor{\log_{10}(v_{ori})}\right \rfloor -1}}
}\right \rceil\cdot 10^{\left \lfloor{\log_{10}(v_{ori})}\right \rfloor -1} \]

This means that all values up to USD 100 are rounded to full 1-dollar values, from 100 and 1~000 to full 10-dollar values, from 1~000 to 10~000 to full 100-dollar values and so on. The expensive strategy reduces the number of buckets further, by scaling them to a magnitude higher values ($v_{exp}$), e.g. to full 10-dollar values between 1 and 100, etc., as follows:
\[v_{exp}= \left \lceil{
\frac{v_{ori}}
{10^{\left \lfloor{\log_{10}(v_{ori})}\right \rfloor }}
}\right \rceil\cdot 10^{\left \lfloor{\log_{10}(v_{ori})}\right \rfloor }\]

The anonymity sets per epoch are shown in Figure~\ref{fig:valueBucketsGen} and Table~\ref{tab:valueBucketsGen}. While some deanonymized payments remain (on average 27.4 (cheap), 1.96 (expensive)), the scaling buckets perform well with anonymity set sizes of mostly at least 10 users - a huge improvement from the original values (nearly 3~000 deanonymized payments) and especially much better than fixed bucket sizes (still 63.7 deanonymized payments for 1~000-dollar buckets). Thus, with scaling buckets there is hope for the value-based anonymity sets, even though they of course are still rather small and should be used with caution.

\begin{table}[htb]
  \begin{center}
    \caption{Lower quartile, median and upper quartile for anonymity set sizes with value buckets under varying frequencies and epochs.}
    \label{tab:valueBucketsGen}
\resizebox{0.95\linewidth}{!}{
\begin{tabular}{|l|c|c|c|}\hline
\diagbox[width=8em]{\textbf{Frequency}}{\textbf{Epoch}}&
  \textbf{12.5 min} & \textbf{25 min} & \textbf{125 min} \\ \hline
 \textbf{high:} 1.0&  1 \textbf{1} 3 & 1 \textbf{1} 4& 1 \textbf{1} 4
 \\
1000.0 &  1 \textbf{2} 8 & 1 \textbf{2} 9& 1 \textbf{2} 10 \\ 
scaled &  12 \textbf{64} 140 & 17 \textbf{117} 269& 30.5 \textbf{468} 1220 \\ 
scaled exp & 79 \textbf{597} 1103 & 94 \textbf{1135.5}  2093.25& 244.5 \textbf{5313.5} 9717.75 \\ 
\hline
\textbf{normal:} 1.0&  1 \textbf{1} 3 & 1 \textbf{1} 3& 1 \textbf{1} 4  \\
1000.0& 1 \textbf{2} 7 & 1 \textbf{2} 7&1 \textbf{2} 9  \\ 
scaled& 5 \textbf{17} 35 & 7 \textbf{32} 66&17 \textbf{124} 289  \\ 
scaled exp& 31 \textbf{142} 250 & 47 \textbf{275} 500&93.5 \textbf{1215.5} 2248  \\ 
\hline
\textbf{low:} 1.0&  1 \textbf{1} 2 & 1 \textbf{1} 3 & 1 \textbf{1} 3  \\
1000.0 &  1 \textbf{2} 7 & 1 \textbf{2} 7& 1 \textbf{2} 8 \\ 
scaled &  4 \textbf{11} 23 & 6 \textbf{21} 43& 13 \textbf{76} 167 \\ 
scaled exp &  22 \textbf{90} 155 & 38 \textbf{175} 308& 65.5 \textbf{704} 1309.5
\\ 
\hline
\end{tabular}}
\end{center}
\end{table}

\section{Simulation for path-based multihop protocols}
We then turn to path-based multihop protocols, and investigate the expected path-based anonymity set size.
General information about sender-receiver relationships and payment paths is too sparse to synthesize this data. 
We hence used real world Ripple data for this study and validate our findings for the other anonymity sets with real world data at the same time.

\subsection{Setting and Data}
\subsubsection{Setting} In addition to the epoch time of Section~\ref{Sec:SimSettingGenerated}, we introduce another type of time slot: the \emph{hop time} as the  delay per single hop.
All payments that are initiated within a hop time slot and are at the same intermediate node are considered to be mixed for the \emph{value and path anonymity set}. As there might be multiple intermediate hops on a path and we like the worst case time to complete a payment to be similar to the one in the other setting, we decide on shorter hop times than epoch times, but also vary them until both times are equal. 

\subsubsection{Data}
We use Ripple payment data from January 2013 until August 2016, provided by~\cite{rippledataset}. 
The dataset contains over 800~000 payments with sender, receiver, value and time stamp. Those payments already have self-transactions filtered out. Further the data contains the payment channel graph and updates for the graph over the same time interval, with graph updates missing for some time spans. As there are on average only 25 payments per hour, we decide to increase time by a factor of 1~000, i.e. each second is interpreted as a millisecond, to work with payment data that is more realistic for current user behavior. Note that the number of payments per time varies greatly from 6.19 to around 70 payments/minute later. We thus use the payments of the last 10 months (October 2015 - August 2016) to avoid effects due to the novelty of the network in earlier years, resulting in 52 payments/minute on average (see Table~\ref{tab:timePerTick} for a comparison to current payment behavior).

To generate a path for each payment, we calculate the shortest path between sender and receiver that has enough capacity to allow for the current payment and update the capacities accordingly, before looking at the next payment. This strategy results in nearly 238 000 successful payments of which nearly 182 000 use at least one intermediate hop and the average number of intermediate hops is 1.23 with a maximum of 10 intermediate hops.

\subsection{Results}
\subsubsection{Anonymity Set Comparison}
Evaluating the strategies above on the Ripple data without paths, we observe that the anonymity sets for active senders and scaled value buckets for an epoch of 25 minutes are much smaller than for the generated data above (cmp. the first three box plots of Fig.~\ref{fig:overview2-25Ripple}).

With regards to the path-based approaches we want to ensure a total processing time that roughly equals 25 minutes or less for each payment, as above. 
Given the maximum path length of 10 hops in our experiment, we decided to set the hop time to 2 minutes such that also payments with the maximum path length are completed on time. Despite this strict choice, the path-based anonymity sets (ignoring the payment values) are surprisingly high, even though there are only 1.23 intermediate nodes per payment path on average (cmp. box plots 4 and 5 in Fig.~\ref{fig:overview2-25Ripple}). 
We gather that many payments are using few, popular intermediate nodes, traversed by many simultaneous transactions. 
Indeed, depending on the hop time, as little as 12-18 nodes are needed to be honest in total, to have at least one honest node on every path in the given data set. 

Testing for the worst case and choosing the least suitable node on each path to be honest, this number increased to  150 different honest nodes.
Note that if there would always be exactly one intermediate node on the path the min and max path-based anonymity sets would be equal. That there is often only one intermediate hop is likely the reason for the good performance even of the min path-based anonymity sets.

Note that, theoretically, the maximum path-based anonymity could yield higher anonymity sets than counting simultaneous active senders: The epochs (used for the active senders) use a fixed time window (e.g. 25 minutes), but the time window in which payments contribute to the maximum path based anonymity sets might exceed the time a single payment needs (e.g. 25 minutes). The reason is that a payment at the very end of its processing (say e.g. at its minute 20) may still mix with a payment that has just begun processing. In this way both; payments starting up to 25 minutes before the payment in question and those starting 25 minutes after it, potentially contribute to the payment's anonymity set. This effect is even amplified by the transitivity of the maximum path based anonymity set. However as the average path length is short in our data, we do not observe such effects in Fig.\ \ref{fig:overview2-25Ripple}.

Let's consider both the path and value, as required for the weakest protection version under investigation, next. This naturally reduces the anonymity set size drastically, as seen above.
Indeed, taking the worst case node choice for each path, we observe that not even the expensive scaling bucket strategy can effectively save the situation (cmp. Fig. \ref{fig:overview2-25Ripple}, 'scale exp + path min'). Only a very small minority of transactions is effectively anonymized in this worst case scenario.

\subsubsection{The Effect of the Hop Times}
\begin{figure}
     \centering
     \begin{subfigure}[b]{0.95\columnwidth}
         \centering
\includegraphics[width=0.95\textwidth]{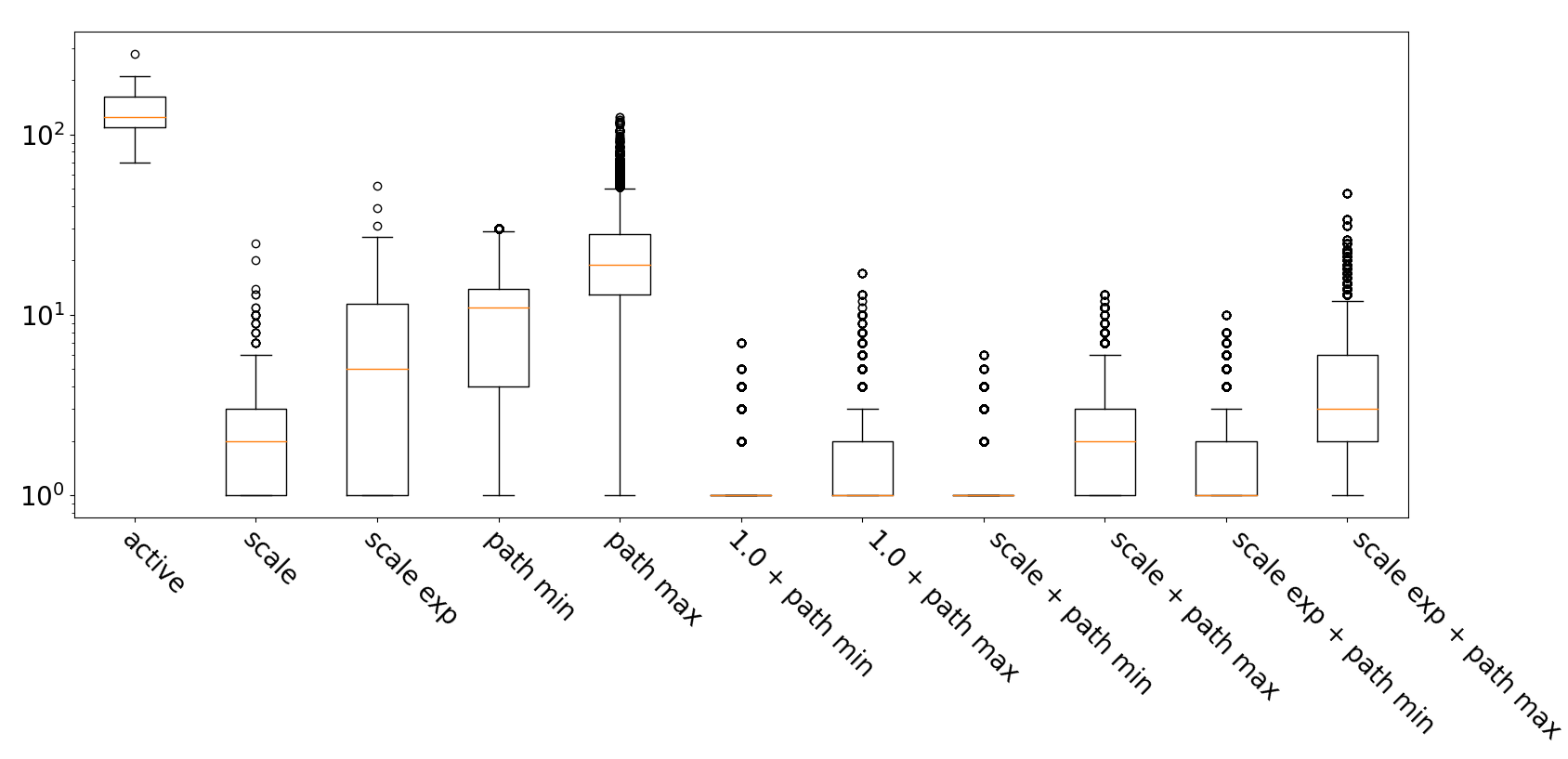}
\caption{Hoptime of 2 minutes. The path-based sets are performing surprisingly well due to the existence of few, very popular nodes at which the transactions get mixed. Leaking values however limits the anonymity sets drastically. }
\label{fig:overview2-25Ripple}
     \end{subfigure}
    \begin{subfigure}[b]{0.95\columnwidth}
         \centering
\includegraphics[width=0.95\textwidth]{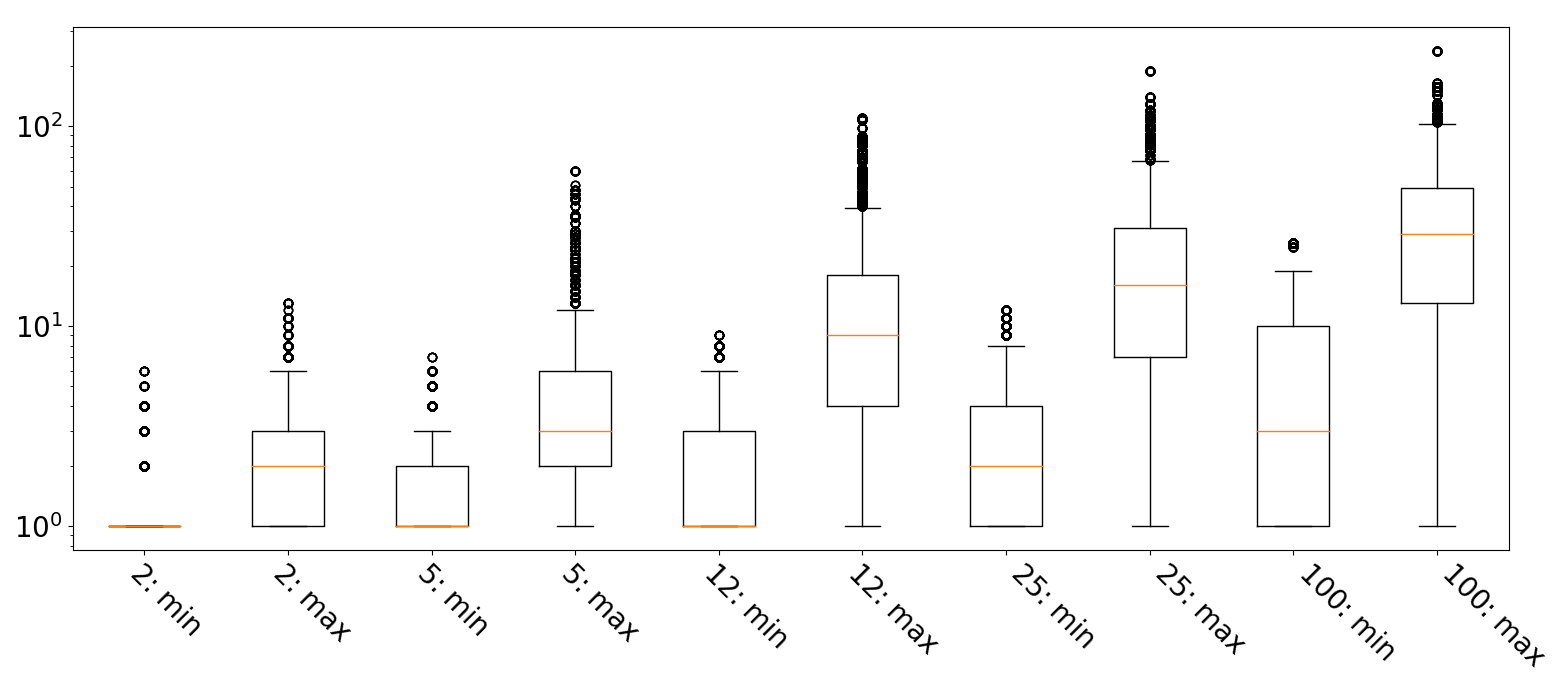} 
\caption{Path anonymity set sizes for the scaled buckets with increasing hop times. As expected with higher hoptime, i.e. latency, the anonymity set sizes grow, however for the minimum anonymity set sizes the numbers are still very small.}
\label{fig:hopEvol}
     \end{subfigure}
    \caption{Ripple Data Anonymity Set Comparison with epochtime of 25 minutes}
        \label{fig:IllustrationsPath}
\end{figure}

Finally, in Figure~\ref{fig:hopEvol} we further investigate the effect of varying the hop times. The anonymity sets grow as expected if higher latency is accepted. However even with tremendous latency (up to 16.67 hours for a 10 hop path) the average size of the minimum path based anonymity sets remains as low as 3.

\section{Summary and Conclusion}\label{sec:conclusion}

In this paper, we demonstrated the practical implications of typical anonymity design decisions for \PPPSystems . Estimating the sizes of corresponding anonymity sets, 
we observed that the expected anonymity set sizes given realistic assumptions is unacceptably low for some of the decisions. We hence investigated suitable strategies to improve the situation. Our results show that delaying transactions may yield low benefit. 
Increasing payment values to rounded classes on the other hand does not impose prohibitive additional cost with scaled buckets, and helps to get closer to acceptable anonymity set sizes.
Looking ahead our simulation study results can be applied to improve existing \PPPSystems .

\bibliographystyle{IEEEtranS}
\bibliography{articles}

\appendix

\subsection{Ignoring Loops}\label{app:loops}
\begin{figure}
\centering
\includegraphics[width=0.49\textwidth]{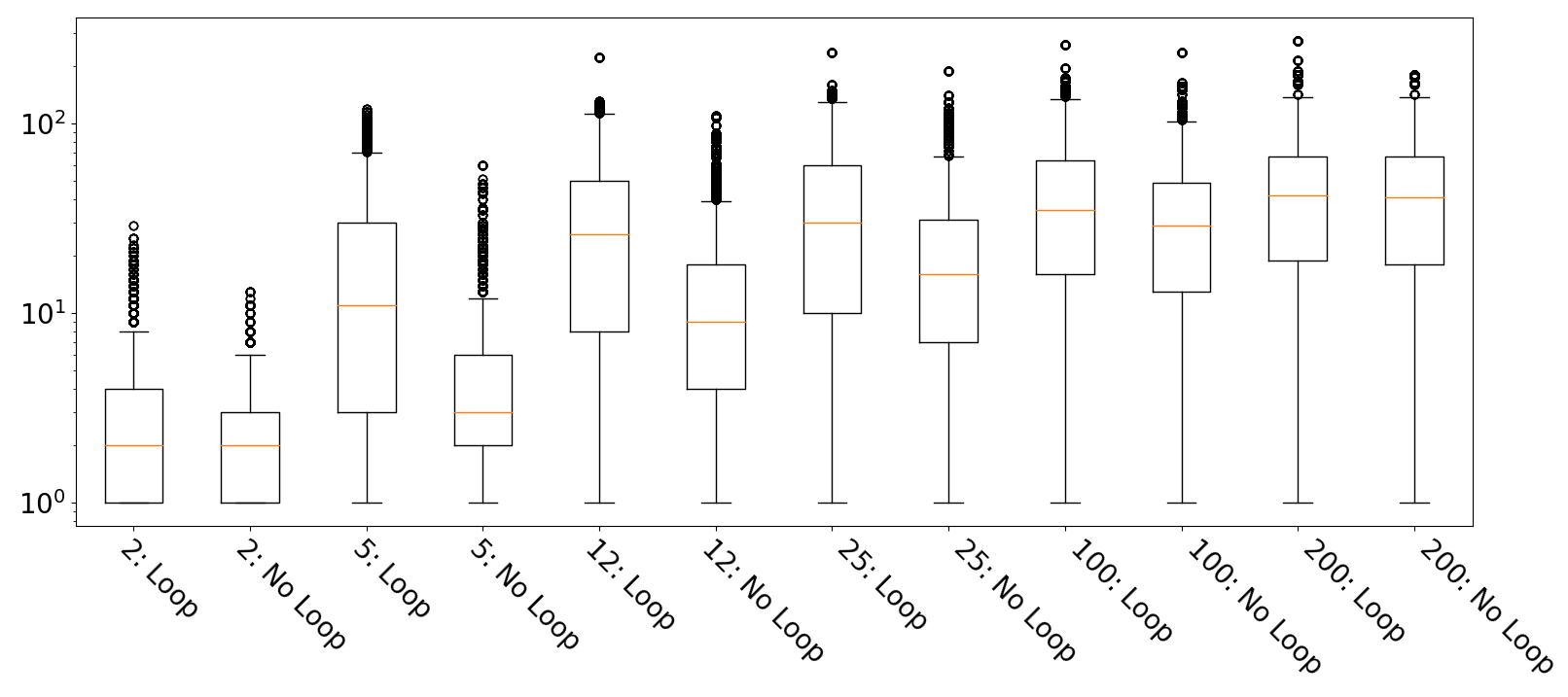}
\caption{Effects of allowing loops illustrated with the maximum path anonymity set sizes of varying hoptime for the scaled buckets. The differences in the anonymity set sizes caused by allowing loops are smaller for very short or long hop times. However the increase when including loops is never high enough to justify introducing loops to the routing as improvement mechanism.}
\label{fig:loops}
\end{figure}

So far we assumed a careful adversary that knows the routing strategy and ignores paths with loops for the maximum anonymity sets. However, to see how much influence loops have on the sets, we calculated the maximum anonymity sets that also allow for loops on the path and show the difference in Figure~\ref{fig:loops}. For short hop times the difference is not high as generally not many payments are mixed and hence not many loops are constructed transitively. For very large hop times, the loops also do not change the anonymity set much as there are not many different time slots in which effects of the transitivity could occur. However, for intermediately long hop times a difference can be seen. As this however not only includes paths with one, but also we multiple hops and as allowing for loops in the paths supports congestion attacks~\cite{evans2009practical}, we do not deem including loops to the paths as a useful countermeasure, but suggest increased hop times instead.

\end{document}